# Symmetric boundary knot method

W. Chen[*]

Simula Research Laboratory, P. O. Box. 134, 1325 Lysaker, Norway

E-mail: wenc@ifi.uio.no

**Abstract**

The boundary knot method (BKM) is a recent boundary-type radial basis function (RBF) collocation scheme for general PDE's. Like the method of fundamental solution (MFS), the RBF is employed to approximate the inhomogeneous terms via the dual reciprocity principle. Unlike the MFS, the method uses a non-singular general solution instead of a singular fundamental solution to evaluate the homogeneous solution so as to circumvent the controversial artificial boundary outside the physical domain. The BKM is meshfree, super-convergent, integration-free, very easy to learn and program. The original BKM, however, loses symmetricity in the presence of mixed boundary. In this study, by analogy with Fasshauer's Hermite RBF interpolation, we developed a symmetric BKM scheme. The accuracy and efficiency of the symmetric BKM are also numerically validated in some 2-D and 3-D Helmholtz and diffusion-reaction problems under complicated geometries.

[*] This research is supported by Norwegian Research Council.

1.Introduction

In recent years much effort has been devoted to developing a variety of meshfree schemes for the numerical solution of partial differential equations (PDE's). The driving force behind the scene is that the mesh-based methods such as the standard FEM and BEM often require prohibitive computational effort to mesh or remesh in handling high-dimensional, moving boundary, and complex-shaped boundary problems. Many of the meshfree techniques available now are based on using the moving least square (MLS) strategy. In most cases, a shadow element is still necessary for numerical integration rather than for function interpolation. Therefore, these methods are not truly meshfree.

Exceptionally, the methods based on the radial basis function (RBF) are inherently meshfree due to the fact that the RBF method does not employ the MLS at all and uses the one-dimensional distance variable irrespective of the dimensionality of the problems. Therefore, the RBF methods are independent of dimensionality and complexity of geometry. Nardini and Brebbia [1] in 1982 have actually applied the RBF concept to develop the currently popular dual reciprocity BEM (DR-BEM) without the notion of "RBF" and the use of then the related advances in multivariate scattered data processing. Only after Kansa's pioneering work [2] in 1990, the research on the RBF method for PDE's has become very active.

Among the existing RBF schemes, the so-called Kansa's method is a domain-type collocation technique, while the method of fundamental solution (MFS) (regular BEM versus singular BEM) is a typical boundary-type RBF collocation methodology. The MFS outperforms the standard BEM in terms of integration-free, super-convergent, easy-to-use, and meshfree merits [3]. The main drawback of the MFS is the use of fictitious boundaries outside the physical domain. The arbitrariness in the determination of the artificial boundary introduces such troublesome issues as stability and accuracy in dealing with complicated geometry systems and undercuts the efficiency of the MFS to practical engineering problems [4,5].

Instead of using a singular fundamental solution, Chen and Tanaka [6,7] exploited a nonsingular general solution to the approximation of the homogeneous solution and removed the controversial artificial boundary in the MFS. The method is called the boundary knot method (BKM). Like the MFS and DR-BEM, the BKM also employs the dual reciprocity method to approximate the particular solution. The method is truly meshfree, integration-free, very easy to learn and program. Some preliminary numerical experiments [6,7] show that the BKM can produce excellent results with a relatively small number of knots for various linear and nonlinear problems. Chen and Hon [9] also numerically illustrated the spectral convergence of the BKM. Spectral convergence here means that the approximation order depends only on the smoothness of the physical solution we approximate. Golberg and Chen [3] also observe that the MFS has spectral convergence.

Like other numerical techniques, the BKM also has its own deficiencies. Besides notably severely ill-conditioned full interpolation matrix [10], the existing BKM loses symmetricity when applied to problems with mixed boundary conditions. In terms of efficiency and stability, symmetricity is widely considered a merit in the numerical solution of PDE's [11]. For example, memory requirements are reduced by a half. By analogy with Fasshauer's Hermite RBF interpolation [12], the purpose of this study is to develop a novel BKM scheme, which holds symmetric interpolation matrix merit for mixed boundary problems without loss of any advantages of the original BKM. It is also worth pointing out that in this study we use the higher order general solution as the RBF's with the dual reciprocity method to evaluate the particular solution. The present symmetric BKM is tested on some 2D and 3D Helmholtz and diffusion-reaction problems under complicated geometry. The paper is concluded with some remarks.

## 2. Symmetric boundary knot method

To clearly illustrate our idea, consider the following example without loss of generality

$$\Re\{u\} = f(x), \ x \in \Omega, \tag{1}$$

$$u(x) = R(x), \ x \in S_u, \tag{2a}$$

$$\frac{\partial u(x)}{\partial n} = N(x), x \in S_T, \tag{2b}$$

where $\Re$ is a linear differential operator, $x$ is the multi-dimensional independent variable, and $n$ is the unit outward normal. The solution of Eq. (1) can be expressed as

$$u = u_h + u_p, \tag{3}$$

where $u_h$ and $u_p$ are the homogeneous and particular solutions, respectively. The latter satisfies

$$\Re\{u_p\} = f(x), \tag{4}$$

but does not necessarily satisfy the boundary conditions.

To evaluate the particular solution, the inhomogeneous term is approximated first by

$$f(x) \cong \sum_{j=1}^{N+L} \lambda_j \varphi(r_j), \tag{5}$$

where $\lambda_j$ are the unknown coefficients. $N$ and $L$ are respectively the numbers of knots inside the domain and on the boundary. $r_j = \|x - x_j\|$ represents the Euclidean distance norm, and $\varphi$ is the radial basis function. Hon and Chen [10] found that inner knots are indispensable to guarantee the stability and accuracy in the DRM and RBF evaluation of the particular solutions. The argument given in [13] that except for improving the solution accuracy, the interior knots are not necessary in the DR-BEM is therefore in doubt.

By forcing the approximation representation (5) to exactly satisfy Eq. (4) at all knots, we can uniquely determine

$$\lambda = A_\varphi^{-1}\{f(x_i)\}, \tag{6}$$

where $A_\varphi$ is the non-singular RBF interpolation matrix. Finally, we can get particular solutions at any point by summing the localized approximate particular solutions

$$u_p = \sum_{j=1}^{N+L} \lambda_j \phi(\|x - x_j\|), \tag{7}$$

where the RBF $\phi$ is related to the RBF $\varphi$ through the operator $\Re$. In this study, we use the second-order general solution of the operator $\Re$ as the RBF $\varphi$, and then the RBF $\phi$ is the corresponding first-order general solution. Some details of the high-order general solution will be elaborated in later section 3. Substituting Eq. (6) into Eq. (7) yields

$$u_p = \Phi A_\varphi^{-1}\{f(x_i)\}, \tag{8}$$

where $\Phi$ is a known matrix comprised of $\phi(r_{ij})$.

On the other hand, the homogeneous solution $u_h$ has to satisfy both the governing equation and the boundary conditions, i.e.,

$$\Re\{u_h\} = 0, \qquad x \in \Omega, \tag{9}$$

$$u_h(x) = R(x) - u_p(x), \qquad x \in S_u, \tag{10a}$$

$$\frac{\partial u_h(x)}{\partial n} = N(x) - \frac{\partial u_p(x)}{\partial n}. \qquad x \in S_T, \qquad (10b)$$

Unlike the DR-BEM [1,13] and MFS [3] using the singular fundamental solution, the BKM [6-8,10] approximates $u_h$ by means of the nonsingular general solution, namely,

$$u_h(x) = \sum_{k=1}^{L} \alpha_k u^{\#}(r_k), \qquad (11)$$

where $k$ is the index of source points on boundary; $\alpha_k$ are the desired coefficients; $u^{\#}$ is the nonsingular general solution of operator $\mathcal{R}$. Refs [7,14] list the general solutions of some often-used differential operators. Collocating Eqs. (10a,b) at all boundary and interior knots in terms of representation (11), we have the unsymmetric BKM scheme:

$$\sum_{k=1}^{L} \alpha_k u^{\#}(r_{ik}) = D(x_i) - u_p(x_i), \qquad x \in \Omega, \qquad (12a)$$

$$\sum_{k=1}^{L} \alpha_k \frac{\partial u^{\#}(r_{jk})}{\partial n} = N(x_j) - \frac{\partial u_p(x_j)}{\partial n}, \qquad x \in S_u, \qquad (12b)$$

$$\sum_{k=1}^{L} \alpha_k u^{\#}(r_{lk}) = u_l - u_p(x_l),\ l = 1,\ldots,N, \qquad x \in S_T, \qquad (12c)$$

where $i, j,$ and $l$ indicate the response knots respectively located on the boundary $S_u$, $S_\Gamma$, and inside the domain $\Omega$. Substituting the approximate particular solution (8) into Eqs. (12a,b,c), we can solve the above simultaneous algebraic equations. After this, we can employ the obtained expansion coefficients $\alpha_k$ and inner knot solutions $u_l$ to calculate the BKM solution at any knot.

The mixed Dirichlet and Neumann boundary conditions lead to the unsymmetric BKM formulations (12a,b). For the domain-type Kansa's method [2], it is even worse that the uniform boundary condition simply destroys the symmetricity. In contrast, it is obvious that if the only one type of boundary condition was involved, the above BKM formulations would be symmetric. Fasshauer [12] recently presented a Hermite interpolation remedy to eliminate the unsymmetric drawback of the Kansa's method.

Following Fasshauer's idea, we modify the naive BKM approximate expression (11) for the homogeneous solution $u_h$ as

$$u_h(x) = \sum_{s=1}^{L_D} a_s u^\#(r_s) - \sum_{s=L_D+1}^{L_D+L_N} a_s \frac{\partial u^\#(r_s)}{\partial n_s}, \qquad (13)$$

where $n_s$ is the unit outward normal at the boundary knot $s$, and $L_D$ and $L_N$ are respectively the numbers of knots at the Dirichlet and Neumann boundary surfaces. The minus sign associated with the second term is due to the fact that the Neumann condition of the first order derivative is not self-adjoint. In terms of expression (13), the collocation analogue equations (12a,b,c) are rewritten as

$$\sum_{s=1}^{L_D} a_s u^\#(r_{is}) - \sum_{s=L_D+1}^{L_D+L_N} a_s \frac{\partial u^\#(r_{is})}{\partial n_s} = R(x_i) - u_p(x_i), \qquad x \in \Omega, \qquad (14a)$$

$$\sum_{s=1}^{L_D} a_s \frac{\partial u^\#(r_{js})}{\partial n_j} - \sum_{s=L_D+1}^{L_D+L_N} a_s \frac{\partial^2 u^\#(r_{js})}{\partial n_{sj}^2} = N(x_j) - \frac{\partial u_p(x_j)}{\partial n_j}, \qquad x \in S_u, \qquad (14b)$$

$$\sum_{s=1}^{L_D} a_s u^\#(r_{ls}) - \sum_{s=L_D+1}^{L_D+L_N} a_s \frac{\partial u^\#(r_{ls})}{\partial n_s} = u_l - u_p(x_l), \qquad x \in S_T, \qquad (14c)$$

where $n_j$ is the unit outward normal at the boundary knot $j$. The system matrix of the above equations is symmetric for mixed boundary problems provided that the operator

ℛ{} involves only even-order derivatives. It is stressed here that the MFS could not produce the symmetric interpolation matrix in any way.

The unsymmetric and symmetric BKM schemes use the expansion coefficients rather than the direct physical variable in the approximation of the boundary value. Therefore, such a BKM is called the indirect BKM. Chen et al. [8] also gave the direct BKM with the physical variable as the basic variable.

## 3. Numerical results and discussions

The 2-D and 3-D irregular geometries tested are illustrated in Figs. 1 and 2, which respectively include an elliptical cutout and a peanut cavity. The parametric surface of the peanut cavity is described by [15]

$$r(\theta,\phi) = R(\theta)\cos(\theta)\vec{i} + R(\theta)\sin(\theta)\cos(\phi)\vec{j} + R(\theta)\sin(\theta)\sin(\phi)\vec{k}, \qquad (15)$$

where $\theta \in [0,\pi)$, $\phi \in [0,2\pi)$. For more details see Fig. 3.

Except for the specified Neumann boundary conditions shown in Fig. 1 and on the $x=0$ surface in Fig. 2, the boundary conditions are all of the Dirichlet type. The BKM employs 15 inner knots for the 2-D inhomogeneous cases as shown by small crosses in Fig. 1. Equally spaced knots were applied on the boundary except on the peanut cavity surface where the random knots were employed. The $L_2$ norms of relative errors were calculated with the BKM solutions at 460 knots for the 2-D, and 1012 knots for the 3-D problems. The absolute error is taken as the relative error if the absolute value of the solution is less than 0.001.

The Helmholtz equation is given by

$$\nabla^2 u + \gamma^2 u = f(x), \qquad (16)$$

The exact solutions of the tested cases are

$$u = x^2 \sin x \cos y \qquad (17)$$

for the 2-D inhomogeneous Helmholtz problem ($\gamma = \sqrt{2}$), and

$$u = \sin x \cos y \cos z \qquad (18)$$

for the 3-D homogeneous Helmholtz problem ($\gamma = \sqrt{3}$).

The diffusion-reaction equation is given by

$$\nabla^2 u - \tau^2 u = q(x), \qquad (19)$$

where $\tau$ is the Thiele parameter. The exact solution is

$$u = e^{-d(x+y)} \qquad (20)$$

for the tested 2-D inhomogeneous diffusion-reaction equation ($\tau = d\sqrt{2}$). The corresponding inhomogeneous functions $f(x)$, $q(x)$ and boundary conditions can be derived accordingly.

In this study, we employ the PDE-dependent second order general solution as the RBF with the dual reciprocity method (i.e. Eq. (5)) to approximate the particular solution. The reason behind this choice is the same as that Golberg and Chen [3] recommend the use of the TPS for the particular solution of Laplace problems. The TPS is understood to be the simplified form of the high-order fundamental solutions of high-order Laplace equations [8]. In the Helmholtz equation case, the RBF $\varphi$ in Eq. (5) is an oscillatory first-order general solution. We think that this is an advantage provided that the external forcing term $f(x)$ is also oscillatory, which happens to occur in the given numerical cases. In addition, the Helmholtz problem, often related to vibration and acoustic wave, in essence is also oscillatory. Thus, the present choice has explicit physical grounds.

Itagaki [16] gives the higher-order fundamental solutions of the 2-D and 3-D Helmholtz and diffusion-reaction Helmholtz operators. We further generalize his results to any dimensions via computer algebra package "Maple", i.e.,

$$u_m^*(r) = A_m (\gamma r)^{-n/2+1+m} Y_{n/2-1+m}(\gamma r), \qquad (21)$$

$$u_m^*(r) = B_m (\tau r)^{-n/2+1+m} K_{n/2-1+m}(\tau r), \qquad (22)$$

where $A_m = A_{m-1}/(2m\gamma^2)$, $B_m = B_{m-1}/(2m\gamma^2)$, $A_0$ and $B_0$ are dimension-dependent constants; $n$ is the dimension of the problem; $m$ denotes order of the general solution; $Y$ and $K$ respectively represent the Bessel and modified Bessel function of the second kind, both of which have a singularity at the origin. Similarly, we derived the higher-order general solutions of Helmholtz and diffusion-reaction operators, i.e.,

$$u_m^\#(r) = A_m (\gamma r)^{-n/2+1+m} J_{n/2-1+m}(\gamma r), \qquad (23)$$

$$u_m^\#(r) = B_m (\tau r)^{-n/2+1+m} I_{n/2-1+m}(\tau r), \qquad (24)$$

where $J$ and $I$ respectively represent the Bessel and modified Bessel function of the first kind. It is noted that the general solution for higher-dimensional problems can usually be significantly simplified. For instance, the zero- and second-order general solutions of the 3-D Helmholtz operator are respectively

$$u_0^\#(r) = \frac{A_0 \sqrt{2}}{\sqrt{\pi}} \frac{\sin(\gamma r)}{\gamma r}, \qquad (25)$$

$$u_2^\#(r) = -\frac{A_2\sqrt{2}}{\sqrt{\pi}} \frac{r^2 \sin(\gamma r) - 3\sin(\gamma r) + 3r\cos(\gamma r)}{\gamma r}. \tag{26}$$

Note that the above zero-order general solution is in fact the known Sinc function. The zero- and second-order general solutions of the 3-D diffusion-reaction operator are respectively

$$u_0^\#(r) = \frac{B_0\sqrt{2}}{\sqrt{\pi}} \frac{\sinh(\tau r)}{\tau r}, \tag{27}$$

$$u_2^\#(r) = \frac{B_2\sqrt{2}}{\sqrt{\pi}} \frac{r^2 \sinh(\tau r) + 3\sinh(\tau r) - 3r\cosh(\tau r)}{\tau r}, \tag{28}$$

where sinh and cosh respectively denote the sine-hyperbolic and cosine-hyperbolic functions.

Tables 1 and 2 respectively display the numerical experimental results of the 2-D inhomogeneous Helmholtz and diffusion-reaction problems, where the numbers inside the parentheses represent the boundary and inside-domain knots. In this study, we used 15 inner knots for all cases. It is found that both unsymmetric and symmetric BKM schemes produce very accurate solutions with a small number of knots. It is also observed that as the Thiele parameter increases, more knots are required to get accurate solutions. Like the DR-BEM and MFS, the BKM is somewhat sensitive to the number and location of inner knots when dealing with inhomogeneous problems. A detailed discussion of this issue is beyond the present study.

The numerical results of the 3D homogeneous Helmholtz case are illustrated in Table 3. No inner knots are required in the BKM for the homogeneous case. The number inside parentheses indicates the number of boundary knots. Relative to the 2-D case, more knots are required to get accurate solutions due not only to the increase in the number of dimensions but also due to a larger domain. It is expected that compared with other numerical techniques, the BKM may become more efficient for higher-dimensional complex-shape problems since the general solutions of high-dimensional operators are simpler and the radial basis function is independent of dimensionality and geometric complexity.

## 4. Remarks

The BKM circumvents the troublesome singular integral inherent in the BEM and is essentially meshfree, integration-free, very easy to learn and program. It is noted that unlike the MFS, the BKM does not require any artificial boundaries outside the physical domain and therefore is very feasible for practical engineering problems with complex-shaped boundary. The study verifies that the BKM can also be a symmetric numerical technique. The given numerical experiments show that the accuracies of the unsymmetric and symmetric BKM schemes have no significant difference. Notwithstanding, the symmetricity is a very favorable property of the numerical technique. In this study, we applied the symmetricity to halve memory requirements. The main focus of this study is that by using a very simple RBF interpolation formula, we derive a symmetric BKM scheme. In contrast, the MFS is non-symmetric in principle.

In particular, we need stress the advantages of the BKM method over the MFS in handling singular problems with nonuniform boundary conditions or sharp corner. Ref. 5 makes the first attempt to apply the MFS to such singular problems successfully only with very tricky problem-dependent placements of boundary knots. The above given examples also involve both the abrupt change of boundary condition and sharp boundary corner. The BKM was found to perform quite well without the sensitivity to the knot placements. Thus, we expect that the BKM may outperform the MFS for a variety of practical engineering computations, where such strong singularities often occur.

It is worth mentioning that as the global interpolation approach, both the BKM and MFS produce a severely ill-conditioned full system matrix if a large number of boundary knots is used, which is very computationally costly. Therefore, developing the corresponding fast algorithms is of vital importance in solving large size problems. In this regard, the fast multipole and domain decomposition techniques [17] seem very promising and are now under study.

**Acknowledgement**

It is gratefully acknowledged that Prof. C.S. Chen kindly provided the surface data of peanut cavity shown in Figs. 2 and 3. The referees' detailed comments enhance the academic quality and readability of this paper.

Fig. 1. Configuration of 2D irregular geometry

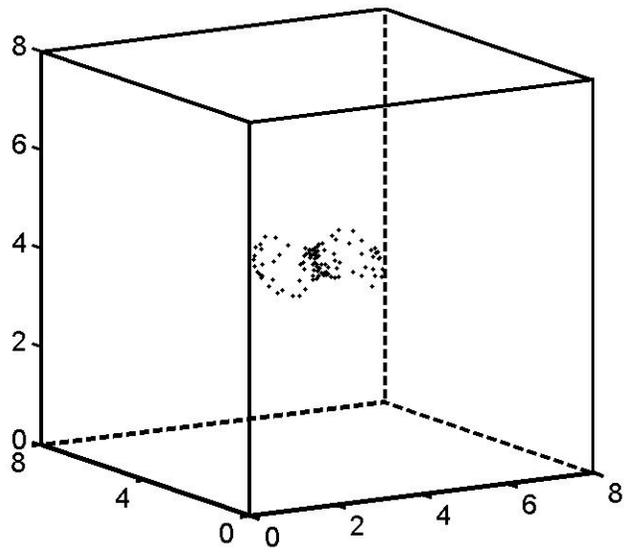

Fig. 2. A cube with small peanut-like cavity

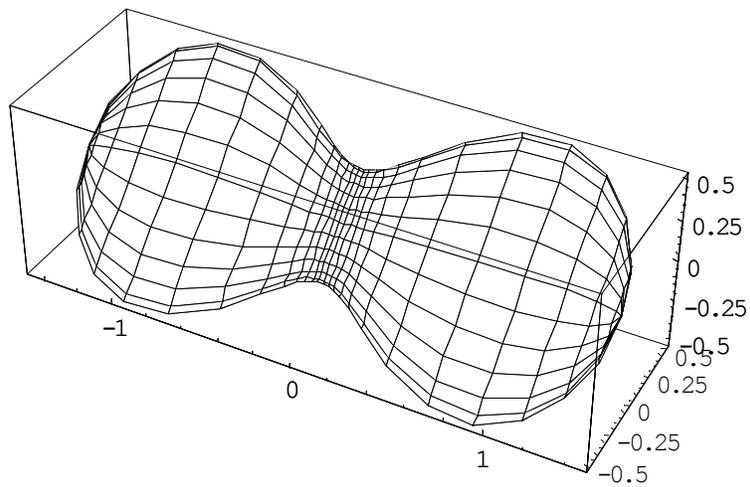

Fig. 3. Parametric surface of the 3D peanut cavity

Table 1. $L_2$ norms of relative errors for 2D inhomogeneous Helmholtz problems by the unsymmetric and symmetric BKM schemes

| BKM (41+15) | BKM (49+15) | SBKM (41+15) | SBKM (49+15) |
|---|---|---|---|
| 5.6e-3 | 4.8e-4 | 1.0e-2 | 1.0e-4 |

Table 2. $L_2$ norms of relative errors for 2D inhomogeneous diffusion-reaction problems by the unsymmetric and symmetric BKM schemes

|  | BKM (25+15) | BKM (33+15) | SBKM (25+15) | SBKM (33+15) |
|---|---|---|---|---|
| $\tau = \sqrt{2}$ | 2.1e-3 | 5.3e-4 | 2.5e-3 | 1.1e-3 |
|  | BKM (49+15) | BKM (57+15) | SBKM (49+15) | SBKM (57+15) |
| $\tau = 5\sqrt{2}$ | 1.3e-2 | 4.5e-4 | 1.6e-2 | 4.2e-4 |

Table 3. $L_2$ norms of relative errors for 3D homogeneous Helmholtz problems by the unsymmetric and symmetric BKM schemes.

| BKM (298) | BKM (376) | SBKM (298) | SBKM (376) |
|---|---|---|---|
| 4.9e-2 | 4.5e-4 | 4.6e-3 | 1.4e-3 |